\documentclass[12pt]{article}

\usepackage{epsf}
\usepackage{amsmath}
\usepackage{graphics}

\renewcommand{\thefootnote}{\#\arabic{footnote}}
\def\AJ{{\it Astron. J.} }

\def\ApJ{{\it Astrophys. J.} }
\def\ApJL{{\it Astrophys. J. Letters} }

\def\ApP{{\it Astropart. Phys.} }
\def\AA{{\it Astron. \& Astroph.} }

\def\MNRAS{{\it Month. Not. Roy. Astr. Soc.} }
\def\Nature{{\it Nature} }

\def\PRB{{\it Phys. Rev.} {\bf B} }
\def\PRD{{\it Phys. Rev.} {\bf D} }
\def\PRL{{\it Phys. Rev. Letters} }

\def\etal{{\it et al.}}

\def\simle{\lower 2pt \hbox {$\buildrel < \over {\scriptstyle \sim }$}}
\def\simge{\lower 2pt \hbox {$\buildrel > \over {\scriptstyle \sim }$}}

\begin{document}

\newcommand{\gtrsim}{ \mathop{}_{\textstyle \sim}^{\textstyle >} }
\newcommand{\lesssim}{ \mathop{}_{\textstyle \sim}^{\textstyle <} }
\newcommand{\rem}[1]{{\bf #1}}
\renewcommand{\thefootnote}{\fnsymbol{footnote}}

\setcounter{footnote}{0}

\begin{titlepage}

\def\thefootnote{\fnsymbol{footnote}}

\begin{center}
\hfill astro-ph/0512188\\
\hfill December 2005\\
\vskip .5in
\bigskip

%
%

\bigskip

{\Large \bf Ultra High Energy Cosmic Rays\\
from Sequestered X Bursts}

\vskip .45in

{\bf Peter L. Biermann}
{\em Max-Planck-Institut f\"{u}r Radioastronomie, Auf dem H\"{u}gel 69,
D-53121 Bonn, Germany}\\
and {\em Department of Physics and Astronomy, University of Bonn, Germany}
\vskip .45in
{\bf Paul H. Frampton}
{\em Perimeter Institute, 31 Caroline Street, Waterloo, ON N2L 2Y5,
Canada}\\
and {\em University of North Carolina, Chapel Hill, NC 27599-3255, USA
\footnote{Permanent address}}

\end{center}

\vskip .4in

\begin{abstract}
Assuming that there is no GZK (Greisen-Zatsepin-Kuzmin) cut-off and that
super-GZK cosmic rays correlate with AGN (Active Galactic Nuclei) at
cosmological distances, it is speculated that a relic superheavy particle
(X) has its lifetime enhanced by sequestration in an extra dimension. This
sequestration is assumed to be partially liberated by proximity of merging
supermassive black holes in an AGN, temporarily but drastically reducing
the lifetime, thus stimulating an X burst. Based on sequestration of the
decay products of X, a speculative explanation of the observed $\gamma/N$
ratio is proposed.
\end{abstract}

\end{titlepage}

\renewcommand{\thepage}{\arabic{page}}

\setcounter{page}{1}

\renewcommand{\thefootnote}{\#\arabic{footnote}}

\newpage

Ultra high energy cosmic rays (UHECR) with primary energy above
the Greisen-Zatsepin-Kuzmin (GZK) energy cut-off have already been
detected, a total of between 10 and 20 times. Although this statement is
not universally accepted, it is believed by a majority of the cosmic-ray
community and will here be assumed along with certain other speculative
assumptions. The source of such UHECR is not well established but here we
assume that at least some of the UHECR originate from radio galaxies, {\it
e.g.} BL Lacs, at cosmological distances of more than 50 Mpc. BL Lac
type sources are a subset of those radio galaxies whose relativistic jet
happens to point at Earth, and so their emission is relativistically
boosted. This boosting provides an enormous selection effect for
detection, and so at 5 GHz radio frequency, for instance, half of all
radio sources are such cases, relativistic jets pointing at Earth - and
demonstrating this by a flat radio spectrum.

There have been a number of hints, that the arrival directions of some
ultra high energy cosmic ray events correlate with the direction to
active galactic nuclei (AGN) (e.g.
\cite{B95,B-MD98,BF98,DTT00,TT01,MB04}), often with active galactic
nuclei at distances much farther than allowed for simple protons or
neutrons due to their interaction with the cosmological microwave
background (for a broad and deep review, see
\cite{St04}). The problem we propose to solve here is how a particle can
survive the microwave background interaction, and yet be correlated with
an active galactic nucleus.

In particular, we shall investigate the possibility introduced
in \cite{CDF} that the longevity of a precursor superheavy particle (X),
along with important properties of the decay of X, are due to sequestering
in a 5th dimension. Some other issues in using superheavy
progenitors for UHECRs are discussed in \cite{Kusenko}.

There may be more than one extra dimension but our mechanism is
adequately illustrated by a 5-dimensional spacetime with coordinates $(t,
x_1, x_2, x_3, y)$ where the space is warped with 3-spatial metric
$|g_{ij}| \sim e^{-Cy}$. The parameter $C$ is of dimension mass and we
may write $C = \alpha M_{string}$ with $M_{string}$ the string scale and
$\alpha$ a parameter of order one to be discussed further below. All
non-gravitational interactions are sequestered on a 3-brane, the real
world brane, which has a thickness extending from $y = 0$ to $y = y_0$.
Gravity, by contrast, is unsequestered and fills the entire 5-spacetime.
By sequestering of X, its decay products, and the standard model particles
at different locations within the real world brane thickness, we shall
construct a model consistent with the observations and the assumption that
the UHECR sources are beyond the GZK mean free path of 50 Mpc.

In order to provide the requisite kinetic energy observed for the
UHECR, the mass of X as top-down progenitor is assumed to be $M_X \sim
10^{14}$ GeV. The lifetime of X in a normal region of approximately flat
spacetime is required \cite{CDF} to be $\tau_X \sim 10^{24}$ s; by
contrast, the typical lifetime for such a particle in particle
phenomenology would be the $\sim 10^{-24}$ s time scale characteristic of
strong interactions. We follow the proposal of \cite{CDF} that the
enormous suppression factor of $\sim 10^{-48}$ arises from sequestering of
the wave function of X relative to its decay products. This is easily
accomplished as the gaussian suppression requires, for $10^{-48} \sim {\rm
exp} (- 110) \simeq {\rm exp} (-[10.5]^2)$, only a separation of the two
wave functions respectively for X and its decay products of some ten times
the characteristic length scale in the brane, typically the inverse of the
string scale, for example $\Delta y \simeq 10^{-35}$ m. Thus this
separation may be as small as $(\Delta y)_{XY} \sim 10^{-34}$ m, but is
correspondingly larger if the string energy scale is reduced.

We shall assume the dominant decay of X is $X \rightarrow Y$ where
$Y \equiv (\bar{{\cal Y}}{\cal Y}$) is a bound state of heavy quarks
${\cal Y}$ bound by QCD into a color singlet hadron $Y$. This hadron Y
exemplifies the ``uhecron" discussed in \cite{AFK}.

Two scenarios must be considered, of which we shall rapidly discard the
first:

\begin{itemize}
\item {\bf Scenario A} Y is absolutely stable and is itself the UHECR
primary.
\item {\bf Scenario B} Y decays into products including nucleons which
act as the UHECR primaries.
\end{itemize}

Before discriminating between these two scenarios for Y, let us
introduce the principal idea concerning the X lifetime and the concept of
an ``X burst". Above, we asserted that the lifetime $\tau_X$ is $\tau_X
\sim 10^{+24}$ s in flat spacetime.

The sources of the UHECR primaries are assumed to be correlated with AGN
in radio galaxies at distances beyond the GZK mean free path,
specifically those radio sources, whose relativistic jet is pointed as
Earth, often also referred to as BL Lac objects. Such AGN in BL Lacs are
associated with mergers of supermassive black holes when the final merger
can take place ``quickly" meaning within, say, one to ten years in the
observer frame at Earth.

Because the gravitational field in 4-spacetime is distorted by the
proximity of the black hole merger, the 5th dimension will likewise be
distorted because of the effects of warping. We shall argue that the
separation $(\Delta y)_{XY}$ can be reduced thereby relative to its value
in flat 4-spacetime by as much as a factor $1/\sqrt{2}$ for distances less
than about ten times the Schwarzschild radii of the black holes. This then
implies the X decay is suppressed by only $\sim 10^{-24}$ rather than by
$\sim 10^{-48}$ and consequently the decay lifetime decreases from
$\tau_X \sim 10^{+24}$ s to $\tau_X^{'} \sim 1$ s whereupon all the X
particles within such proximity of the AGN will decay in an ''X burst". We
shall estimate the consequent flux of UHECRs below.

First it behooves us to justify the effect of AGN warping on the X
lifetime. We can make a qualitative justification, which is sufficient to
illustrate the plausibility, as follows. Bearing in mind the warping
factor $e^{-\alpha M_{string}y}$ in the spatial metric the $|g_{ij}|$
becomes

\begin{eqnarray}
|g_{ij}| & = & \left( 1 - \frac{2 G M c^2}{r} \right)^{-1} {\rm
exp}(-\alpha M_{string} y) \nonumber \\ & = & (1 - r_S/r)^{-1} {\rm exp}(-\alpha
M_{string} y)
\label{warpedschwarzschild}
\end{eqnarray}

Based on this expression (\ref{warpedschwarzschild}) where $r_S$ is
the Schwarzschild radius consider the change from $r = 2 r_S$ to $r = 10
r_S$ where the unwarped factor $(1 - r_S/r)^{-1}$ varies from 2.00 to 1.11
(it is unity at $r = \infty$) so the suggested change in $(\Delta y)_{XY}$
of $3 M_{string}^{-1}$ will occur if we choose $\alpha = [{\rm ln}
(2.00/1.11)]/3 = 0.20$ in the space metric $g_{ij}$. This is of order one
and suggests such liberation of sequestration can occur near an AGN.

Because the ${\cal Y}$ particles are sequestered closer in the y
direction to X than are any of the standard model particles $\{q, l, \nu;
g, W, Z, \gamma\}$ the decay which dominates is BR$(X \rightarrow
Y(\bar{{\cal Y}}{\cal Y}) = 100\%$ by the mechanism espoused in
\cite{AS,MS}.

The particle $Y^0 = (\bar{{\cal Y}}{\cal Y})$ is a bound state hadron
comprised of the heavy quarks ${\cal Y}$. Because ${\cal Y}$ is
sequestered away from $\gamma$ the cross section for scattering on the CMB
is much smaller than for normal hadrons: $\sigma (\gamma Y) << \sigma
(\gamma N)$ and so the mean free path for ultra high energy Y with $E \sim
10^{23}$ eV through the background radiation can be several Gpc or longer.
The extra dimension thus facilitates avoidance of any GZK cut off.

At this stage, it is necessary to pursue separately the scenarios {\bf A}
and {\bf B}.

In Scenario {\bf A}, Y is stable and is the UHECR primary. According to
\cite{AFK}, there is an upper limit on the mass of such a strongly
interacting primary for the AGASA super-GZK events with $E > 100 \, EeV$
of $M_Y < 50$ GeV. At the same time, experiments at the Tevatron analysed
in \cite{BCG} and \cite{MR} would have discovered a Y particle with any
mass $M_Y < 180 $ GeV. Consequently scenario {\bf A} is strongly
disfavored by the non-observation of $Y$ at the Tevatron.

We are therefore led strongly to prefer scenario {\bf B} where there is
the decay chain $X \rightarrow Y \rightarrow N$.

There is now one final sequestration effect acting on the particle Y,
additional to those allowing the longevity of X and the high BR$(Y
\rightarrow X) \simeq 1$. This is to overcome the problem that Y decay
generically leads to too high a $\gamma/N$ (and $\nu/N$) ratio; for
example, in \cite{BKV} and \cite{PJ}, a $(\gamma + \nu)/({\rm total})$
ratio of decay products as high as 97\% is derived for a top-down
progenitor (without extra dimension(s)). This has a simple physical
explanation that the hadronization produces very many pions which yield
photons and neutrinos. However, such a high percentage is inconsistent
with the AGASA events which are believed to be caused by predominantly
strongly interacting primaries, nucleons or light nuclei. Incidentally,
this is an apparent problem for the Z-burst mechanism \cite{W} since the
well measured branching ratios for Z decay would naively lead to too high
$\gamma/N \sim 10$.

The resolution is that Y is sequestered nearer to $\{c, b, t; g\}$ than to
$\{u, d, s, \nu, l; W, Z, \gamma\}$. Note that the sequestration of $Y$
from $\gamma$ is required also for avoidance of the GZK cut off so this
relates the potentially observed violation of GZK with the observed low
$\gamma/N$ ratio which now can be $\gamma/N < 1$.

Let us then summarize the sequestration sequence across the real world
brane from $y=0$ to $y=y_0$ in flat 4-spacetime. Starting at one side,
say, $y=0$ (it does not matter as the two sides are equivalent) we have
$Y$, then $X$, then $\{q; g\}$, and finally $\{\nu, l; w, Z,
\gamma\}$ at $y=y_0$, localized at quite different y values $y=0, ~~ y_X,
~~ y_{QCD}, ~~ y_0$ across the thickness of the real world brane.

\bigskip

To obtain an upper limit on the UHECR flux from the X-burst mechanism,
let us first assume that the X relic constitutes {\it all} the
nonbaryonic dark matter. As a reference galaxy with a massive black hole
we conservatively adopt a galaxy such as M87, the closest nearby powerful
radio galaxy, demonstrated to be able to produce high energy particles
via Fermi acceleration to about $10^{21}$ eV (\cite{BS87}); similar
active galactic nuclei have been observed out to redshift 6.41
(\cite{FNL01}), far beyond the most active cosmological redshift range of
high galaxy merger rates. We adopt the following parameters for such an
active galaxy: The black hole has $3 \times 10^9 \; {\rm M_{\odot}}$, the
galaxy itself has a dark matter halo of $ 2 \times 10^{13} \; {\rm
M_{\odot}}$, and an outer radius of 300 kpc. Then the Schwarzschild
radius of the black hole is
$r_S \; = \; 3 \times 10^{-4} {\rm pc}$.

With mass $M_X = 10^{14}$ GeV this implies a mean number density $\sim
1/(km)^3$ throughout the Universe. The galactic mass is $\sim \;
3 \times 10^{70}$ GeV and so there will be $\sim 3 \times 10^{56}$ X
particles per galaxy. To be conservative, we adopt the universal dark
matter profile from \cite{NFW}

\begin{equation}
\rho_{DM} \; = \; \frac{\rho_{DM, 0}}{(r/r_0)(1 + r/r_0)^2}
\end{equation}

Other scalings and descriptions of the inner dark matter profile
(\cite{GP04,BM05}) give only weakly different numerical estimates. We
adopt a DM scale of $r_0 \; = \; 3 \; {\rm kpc}$. This integrates to give
the mass inside radius $x = r/r_0$

\begin{equation}
M_r \; = \; 4 \pi r_0^3 \; \rho_{DM, 0} \; \left(-\frac{x}{1+x} + \ln (1 +
x) \right)
\end{equation}

Around the black hole there is an increase in dark matter particles, as
they are swept up in the black hole from low angular momentum orbits; the
low angular momentum orbits are repopulated by gravitational disturbances,
and so we have a density law in this region of $r^{-7/4}$
(\cite{H75,BW76,FR76}); this ``loss cone" refilling starts at about the
Bondi-Hoyle radius $r_B$ (e.g. \cite{GP04})

\begin{equation}
r_{B} \; = \; \frac{2 G M_{BH}}{\sigma^2}
\end{equation}

where the black hole begins to dominate the motions and where we adopt
$\sigma \; = \; 500 \, {\rm km/s}$ for the stellar and DM particle
velocity dispersion. For our putative black hole this is $r_{B} \; = \;
100 \; {\rm pc}$. Matching the universal dark matter density at the
Bondi-Hoyle radius gives for the mass inside 10 Schwarzschild radii then

\begin{equation}
M(10 r_S) \; = \; M_{DM, tot} \; \frac{1}{5} \; (\frac{r_B}{r_0})^2 \;
(\frac{10 \; r_S}{r_B})^{5/4}
\end{equation}

We consider this black hole to be fed via a merger with another galaxy,
which gives us an AGN, as noted above. We have checked the scaling
implied by these relations using our own Galaxy, and determined that
these numbers are low estimates for the DM density, so rather
conservative (e.g., \cite{GP04,BM05}). We emphasize again, that strong
effects in General Relativity are expected within a few units of the
innermost stable orbit, which for a maximally rotating black hole varies
from just half a Schwarzschild radius for corotation, to 4.5
Schwarzschild radii at counter-rotation, and 3 Schwarzschild radii for no
rotation.

The dark matter mass inside 10 Schwarzschild radii $r_S$ is then given by
a fraction of about $6 \times 10^{-10}$ of all dark matter particles. So
an X-burst involves the rapid sequestered decay of $1.5 \times 10^{47}$ X
particles in close proximity of a supermassive black hole merger.

In X particle decay, although the decay spectrum is obviously not
measured a crude but adequate upper limit for the number of Y particles
and their decay hadron products produced near or below $10^{11}$ GeV is
about $\sim 10^{48}$ particles per sequestered X burst. We note that
this corresponds to a degeneration of energy content by a factor of $6 \,
10^{-3}$, since we have 6 particles of $10^{11}$ GeV for one particle of
$10^{14}$ GeV. We will refer to this factor as an efficiency
$\epsilon_{\star}$ below.

To estimate the UHECR flux we assume as typical AGN distance the Hubble
distance of $d_{AGN} = 4.5 \, {\rm Gpc} \; \simeq \; 1.5 \times 10^{23}$
km and one such merger per year. The spherical area at that distance is $4
\pi (d_{AGD})^2 \; = \; 3 \times 10^{47} {\rm km^2}$ and so the maximum
flux at the Earth is seen to be

\begin{equation}
\sim 300 {\rm /km^2 / century }
\end{equation}

\bigskip

We could use here a model for the cosmological evolution, and also a
model for the luminosity function of active black holes, i.e. AGN,
at high redshift (see, e.g., \cite{WB00}). We note that the galaxies which
are the most massive today, have the biggest black holes, and so had the
highest merger rate in the past (see, e.g., \cite{GSW05,RBF05});
considering all the uncertainties inherent in our estimate, our simple
approach should suffice for now.

\bigskip

The observed flux of super-GZK events is about one per square kilometer
per century but we assumed that {\it all} dark matter is made from X
particles and if instead we make a much more reasonable assumption that
just a few percent of dark matter is involved then the flux rate of UHECR
becomes sufficiently close to observation to encourage us to take the X
burst mechanism as a speculative but serious candidate for the new
physics (see, e.g., \cite{BKM05} for a view what dark matter might be).

We can perform several more tests:

First we can estimate the luminosity produced in this mechanism. We
consume all the X-particles within the 10 Schwarzschild radii in our
simple approximation. This consumption is at about the escape speed at
that distance, which is about $c/5$.

\begin{equation}
L_X \; = \; \frac{M(10 \; r_S) \; c^2}{50 \; r_S /c} \; \epsilon_{\star}
\end{equation}

This yields then

\begin{equation}
L_X \; = \; 4 \; 10^{49} \; {\rm erg/s} \; \frac{M_{BH}}{3 \; 10^9 \;
M_{\odot}} \; F
\end{equation}

with

\begin{equation}
F \; = \; \frac{M_{BH}}{3 \; 10^9 \; M_{\odot}} \; (\frac{3 \; {\rm
kpc}}{r_0})^2 \; (\frac{\sigma}{500 \; {\rm km/s}})^{-3/2} \;
\frac{\epsilon_{\star}}{6 \, 10^{-3}}
\end{equation}

The fundamental plane correlations of elliptical galaxies
(\cite{FDD86}) show that this function $F \; \approx \; 1$ is only weakly
dependent on mass of the black hole, or mass of the galaxy, with an
approximate relation of $F \; \sim \; {M_{BH}}^{-3/8}$.

This suggests one more time, that we overestimate the amount of dark
matter contributing in X particles. If we insert for both relationships
a factor of $1/100$, then we finally predict a flux contribution of 1
particle per km$^2$ and per century, and a corresponding luminosity in
high energy cosmic rays of

\begin{equation}
L_X \; \approx \; 4 \; 10^{47} \; {\rm erg/s} \; \frac{M_{BH}}{3 \; 10^9
\; M_{\odot}}.
\end{equation}

It is notoriously difficult to estimate observed flux from single events,
but a very crude estimate can be gotten this way, and it suggests that
the sources are emitting some fraction of the Eddington luminosity, which
happens to be equal to the numbers given above in the last equation: The
Eddington luminosity is that luminosity for which gravitational
attraction and radiative repulsion balance for an electron/proton plasma.
In our context this measure has no physical relevance except that
observationally we happen to know that all well determined sources appear
to obey this limit. Therefore the luminosity above derived from an
entirely different argument is consistent with observations.  

Second we can also estimate the directionality: Since the high energy
cosmic ray events correlate with relativistic jets pointing at us - flat
spectrum radio sources are all such jets, \cite{W+88} - the squeezing of
space around the merger must also produce a directionality in the emitted
Y-particles, with similar angular opening. This directionality is almost certainly the direction of the spin
of the merged black hole; that is the dominant spin in the system. This
squeezing once again enhances the apparent luminosity, and so we could lower the fraction of dark matter associated
with the X particles even further. Otherwise the correlation would
disappear.

More correctly, the alignment of the Y decay products from X decay
with the BL LAC jet axis is expected to result not from a squeezing of 
three normal spatial dimensions but of the additional dimension. The
lifetime of X is expected to be the most shortened 
for decay $X \rightarrow Y$ in the axial direction as a 
qualitative result of the scale in the extra dimension
as in Eq.(\ref{warpedschwarzschild}) tracking the
scale of the ergosphere in the familiar Kerr solution
which the newer five-dimensional solutions \cite{Emparan,Horowitz} generalize. 
In the well known Kerr solution the ergosphere
has its largest scale equatorially, and here the warping of the fifth
dimension is correspondingly weakest. The concept of an
anisotropic lifetime may sound unfamiliar but is not surprising
for a lifetime hypersensitive to the warping factor.
Our discussion is necessarily only qualitative but the
beaming of the Y particles from X bursts along the BL LAC jet 
by this mechanism is quite plausible.

Third we can work out the time spread: The emisson is made with
a time of ${50 \; r_S /c}$, which for a black hole of mass of $3 \; 10^9$
solar masses is about a week. And yet, since the estimated
lifetime of activity of flat spectrum radio sources is believed to be
about $10^8$ years, the arrival time of the decay products of the Y
particle may be spread out, and that can be understood as the time
spread from the decay at various distances along the path to us, which is
given by a fraction of $1/(2 \gamma_Y^2)$ of the travel time from the
source to us. The travel time for a photon at our adopted standard
distance is of order $10^{10}$ years, and so the Y has to have a Lorentz
factor of at least $10$, or a mass of less than $10^{13}$ GeV. If the mass
were larger, then the arrival would not correlate with an activity
episode of an AGN anymore, since the spreading would be longer than the
activity episode lifetime of the AGN. However, the spreading could be
smaller, since we associate only a small fraction of the visible flat
spectrum radio sources with the high energy particle emission. Any
correlation with just a few soures, which can each supply a major fraction
of the flux,  suggests, that in the observer's frame the time spread is of
order 1 - 10 years, and this would suggest that the Y has a mass of order
a few $10^9$ GeV. Obviously, this is a very crude estimate, since we
should really be doing a convolution with cosmological evolution, and
with a distribution function of AGN power levels. With good statistics of
associations with flat spectrum radio sources we might be able to derive
a better estimate for the mass of the Y. But this should point to the
right range.

Fourth, we can obtain a limit on the lifetime of the Y particle in the
observer frame: For particles that decay within a small fraction of the
path to us, the resulting hadronic flux is exponentially suppressed by the
interaction with the microwave background; while for Y particles whose
decay time is much longer than the transit time, the rate of decay at our
distance could be quite low, but this suppression is only with the
inverse of the lifetime; however, this last effect could be compensated by
some of the other effects, which increase the flux, such as beaming (see
above). Basically, the lifetime in its own frame has to be of order
$10^9$ years or some factor of order at most 100 longer.

Last we wish to suggest how to test this proposal, and how
to distinguish it from other suggestions to explain such a possible
correlation between UHECR and AGN: there is basically one other
suggestion, and that is an effect of quantum gravity could shift the
threshold for the interaction between protons and neutrons, in such a
way, that protons would decay, and neutrons would survive in the
interaction with the microwave background (e.g., \cite{CG99,ACP01}). This
would again imply a directional correlation, because the particles
surviving are neutrons; however, the basic injection would be just the
same as for normal radio galaxies (e.g. \cite{BS87,RB93,RSB93}). This
implies that the overall integrated injection spectrum of cosmic rays
from such sources, taking into account the dependence of maximum energy
on power of the source (\cite{BBR05}), is fairly steep. Here in our
model, the injection spectrum would reproduce the decay spectrum of the
Y-particle, and so would be presumably quite a bit flatter.

A corrolary is that dark matter does have this small contribution from
such X-particles.

Therefore, the observational test is the injection spectrum required to
explain those UHECR events, that do correlate with AGN at cosmological
distances.

\bigskip

UHECR ($> 100$ EeV) are exciting as signals of new physics provided
certain correlations which are still uncertain become better established.
Especially, one would want to obtain an improved statistical significance
of the correlations of the directions of the UHECR with the directions of
the radio galaxies with AGN (BL LACs) at distances $>$ 50 Mpc (see, e.g.,
\cite{DTT00}), as first discussed in a physical context in, e.g., \cite
{RB93,B95,BF98}.

The Auger detector, as well as EUSO and OWL, are expected to provide
higher statistics for the UHECR. The first question to answer is whether
such cosmic rays exist? (Fluoresence and ground-based measurements in
AGASA, Hi-Res and preliminary Auger data seem only marginally consistent).
Second question: do the air-shower analyses confirm that primaries are all
or mostly hadronic? Third question: Is there a statistically significant
correlation between UHECR (assuming such exist!) and AGNs at cosmological
distances?

If the answers to all three questions are positive then it does seem that
dramatically new physics will have been discovered as these three facts
cannot be accommodated with known particles and forces.

Extra dimensions are purely speculative at present. They just might
show up at the planned colliders but they may have insufficient energy.
The uniquely high energies of 100 EeV cosmic rays could be the first
opportunity to detect extra dimensions in the early 21st century just as
cosmic rays led to the original discoveries of important elementary
particles like $e^+, \mu, \pi, K$ and others in the first half of the 20th
century.

\bigskip

\newpage

\bigskip
\bigskip
\bigskip
\bigskip

\begin{center}
{\bf Acknowledgements}
\end{center}

\bigskip
We both acknowledge the hospitality of the Aspen Center for Physics and
thank the participants, including those in the cosmic ray workshop, for
useful discussions. The work of P.H.F. is supported in part by the US
Department of Energy under Grant No. DE-FG02-97ER-41036. The work of
P.L.B. is supported through the AUGER theory and membership grant
05 CU 5PD1/2 via DESY/BMBF (Germany). We would like to acknowledge
recent discussions pertaining to the cosmological evolution and the
activity of black holes with R. Juszkiewicz, H. Kang, A. Kusenko, F.
I. Mari\c{s}, Munyaneza, D. Ryu, and Y.P. Wang.


\newpage

\bigskip

\begin{thebibliography}{100}
\bibitem{B95}
Biermann, P.L., in {\it Nucl.Phys. B Suppl.} {\bf 43}, 221 (1995).
%

\bibitem{B-MD98} in Proc. of the NASA Workshop on Observing Giant
Cosmic Ray Air Showers for $E > 10^{20}$ eV Particles from Space',
Eds: John F. Krizmanic, Jonathan F. Ormes, and Robert E.
Streitmatter , in AIP conf. proc. 433, book, p. 22 - 36, 1998

\bibitem{BF98}
P.L. Biermann and G. Farrar, \PRL {\bf 81,} 3579 (1998).
{\tt astro-ph/9806242}.

\bibitem{DTT00}
Dubovsky, S. L., Tinyakov, P. G., Tkachev, I. I., \PRL {\bf 85}, 1154
(2000). {\tt astro-ph/0001317}.

\bibitem{TT01} Tinyakov, P. G., Tkachev, I. I., {\it JETP Lett.}, {\bf
74}, 1 - 5 (2001). {\tt astro-ph/0102101}

\bibitem{MB04}  
Mari\c{s}, I., Biermann, P.L., to be published; see
research report 2001 - 2004 on www.mpifr-bonn.mpg,de/div/theory
%

\bibitem{St04}
Stanev, T., {\it High Energy Cosmic Rays}, Springer (Berlin 2004).

\bibitem{CDF}
J.L. Crooks, J.O. Dunn and P.H. Frampton, \ApJL {\bf 546,} L1
(2001). {\tt astro-ph/0002089}.

\bibitem{Kusenko}
A. Kusenko and V.A. Kuzmin, JETP Lett. {\bf 73,} 443 (2001).
{\tt astro-ph/0012040}.
G. Sigl, Annals Phys. {\bf 303,} 117 (2003). 
{\tt astro-ph/0210049}.
V.K. Dubrovich, D. Fargion and M.Yu. Khlopov, Astropart. Phys. {\bf 22,}
183 (2004). {\tt hep-ph/0312105}.

\bibitem{AFK}
I.F.M. Albuquerque, G.R. Farrar and E.W. Kolb, \PRD {\bf 59,}
015021 (1998). {\tt hep-ph/9805288}.

\bibitem{AS}
N. Arkani-Hamed and M. Schmaltz, \PRD {\bf 61,} 033005 (2000).
{\tt hep-ph/9903417}.

\bibitem{MS}
E.A. Mirabelli and M. Schmaltz, \PRD{\bf 61,} 113011 (2000). {\tt
hep-ph/9912265}.

\bibitem{BCG}
H. Baer, K. Cheung and J.F. Gunion, \PRB {\bf 59,} 075002 (1999).
{\tt hep-ph/9806361}.

\bibitem{MR}
A. Mafi and S. Raby, \PRD {\bf 62,} 035003 (2000). {\tt
hep-ph/9912436}.

\bibitem{BKV}
V. Berezinsky, M. Kachelriess and A. Vilenkin, \PRL {\bf 79,}
4302 (1997). {\tt astro-ph/9708217}.

\bibitem{PJ}
R.J. Protheroe and P.A. Johnson,
Nucl. Phys. Proc. Suppl {\bf 48,} 485 (1996). {\tt astro-ph/9605006}.

\bibitem{W}
T.J. Weiler, \PRL {\bf 49,} 234 (1982).

\bibitem{BS87}
Biermann, P. L., Strittmatter, P. A., \ApJ {\bf 322}, 643 (1987).
%

\bibitem{FNL01}
Fan, X. \etal, \AJ, {\bf 122}, 2833 (2001). {\tt astro-ph/0108063}.
%

\bibitem{NFW}
Navarro, J. F., Frenk, C. S., White, S. D. M., \ApJ {\bf
490}, 493 (1997). {\tt astro-ph/9611107}
%

\bibitem{GP04}
Gnedin, Oleg Y., \& Primack, J. R., \PRL {\bf 93}, 061302 (2004).
{\tt astro-ph/0308385}
%

\bibitem{BM05}
Bertone, G., \& Merritt, D., \PRD {\bf 72}, 103502 (2005). {\tt
astro-ph/0501555}
%

\bibitem{H75}
Hills, J. G., \Nature {\bf 254}, 295 (1975).
%

\bibitem{BW76}
Bahcall, J. N., Wolf, R. A., \ApJ {\bf 209}, 214 (1976).
%

\bibitem{FR76}
Frank, J., Rees, M. J., \MNRAS {\bf 176}, 633 (1976).
%

\bibitem{WB00}
Wang, Y.-P., \& Biermann, P.L., \AA {\bf 356}, 808 (2000). {\tt
astro-ph/0003005}

\bibitem{GSW05}
Gao, L., Springel, V., White, S.D.M., \MNRAS {\bf 363}, L66 - L70 (2005)
%

\bibitem{RBF05}
Reed, D.S., Bower, R., Frenk, C.S. \etal, \MNRAS {\bf 363}, 393 - 404
(2005)
%

\bibitem{BKM05}
Biermann, P.L., in Proc. {\it Carpathian Summer School in Physics 2005},
Ed. L. Trache, Texas A\& M, World Scientific (in press 2005). {\tt
astro-ph/0510024}
%

\bibitem{FDD86} Faber, S.M., \etal, in Proc. of {\it Nearly Normal
Galaxies}, 8th Santa Cruz Workshop, Ed. S.M. Faber, Springer, Berlin, p.
175 - 183 (1987)
%

\bibitem{W+88} Witzel, A., \etal, \AA {\bf 206}, 245 - 252 (1988).
%
\bibitem{Emparan}
R. Emparan and H.S. Reall, Phys. Rev. Lett. {\bf 88,} 101101 (2002).
{\tt hep-th/0110260}.
\bibitem{Horowitz}
For a review, see {\it e.g.} G.T. Horowitz. {\tt gr-qc/0507080}.
\bibitem{CG99}
Coleman, S. \& Glashow, S. L., \PRD {\bf 59}, 116008 (1999). {\tt
hep-ph/9812418}
%

\bibitem{ACP01} Amelino-Camelia, G., Piran, T., \PRD {\bf 64}, 036005
(2001). {\tt astro-ph/0008107}

\bibitem{RB93}
Rachen, J.P., \& Biermann, P.L., \AA {\bf 272}, 161,
(1993). {\tt astro-ph/9301010}
%

\bibitem{RSB93}
Rachen, J.P., Stanev, T., \& Biermann, P.L., \AA {\bf
273}, 377 (1993). {\tt astro-ph/9302005}

\bibitem{BBR05}
Becker, J. K., Biermann, P. L., \& Rhode, W., \ApP {\bf 23}, 355 (2005).
{\tt astro-ph/0502089}
%

\end{thebibliography}
\end{document}